\documentclass[12pt]{article}
\usepackage{graphicx}
\usepackage{endfloat}

\begin{document}
\title{ Simulating decoherence behavior of a system in entangled state using nuclear
magnetic resonance }
\author{\small{} Jingfu Zhang$^{1}$ , Zhiheng Lu$^{1}$,Lu Shan$^{2}$,and Zhiwei
Deng$^{2}$  \\
\small{} $^{1}$Department of Physics,\\
\small{}Beijing Normal University, Beijing,
100875, Peoples' Republic of China\\
\small{} $^{2}$Testing and Analytical Center,\\
\small{}  Beijing Normal University,
 Beijing,100875, Peoples' Republic of China}
\date{}
\maketitle
\begin{center}\bf Abstract\end{center}
\begin{minipage}{120mm}
\hspace{0.5cm} {\small By choosing $^{1}H$ nucleus in Carbon-13
labelled trichloroethylene as one qubit environment, and two
$^{13}C$ nuclei as a two-qubit system, we have simulated quantum
decoherence when the system lies in an entangled state using
nuclear magnetic resonance (NMR). Decoupling technique is used to
trace over the environment degrees of freedom. Experimental
results show agreements with the theoretical predictions. Our
experiment scheme can be generalized to the case that environment
is composed of multiple qubits.}

PACS number(s):03.67
\end{minipage}

%------------------------------------------------------
\vspace{0.5cm}
 Quantum decoherence is a purely quantum-mechanical
effect through which a system loses its coherence behavior by
getting entangled with its environment degrees of freedom [1][2].
A real quantum system always interacts with its surrounding
environment. The time evolution induced by the interaction
introduces entanglement between the system and  environment when
the system initially lies in a superposition of states. When the
state of the system is described by a reduced density matrix
through tracing over the environment degrees of freedom,
decoherence makes the off-diagonal matrix elements approach 0, and
leaves diagonal ones unaltered. Quantum decoherence is thought as
a main obstacle for experimental implementations of quantum
computation and has been widely studied in theory [3]-[5]. Some
theoretical and experimental schemes have also been proposed to
solve the problem of decoherence in order to fully use coherence
in quantum information[6]-[8]. C.J. Myatt. et al observed and
studied decoherence by coupling an ion in a Paul trap to a
reservoir that can be controlled[9][10]. D. G. Cory et al
simulated decoherence by magnetic field gradients [7]. In this
paper, we will simulate decoherence when the system lies in an
entangled state. Although we choose one qubit to simulate the
environment, our scheme can be generalized to the case that the
environment is composed of multiple qubits.

We choose carbon 13-labelled trichloroethylene (TCE) dissolved in
d-chloroform as a sample. TCE's structure is shown in Fig.1. The
two $^{13}$C nuclei constitute the system, and the $^{1}$H nucleus
is viewed as environment. They represent qubits 1, 2, and 3,
respectively. The influence of the other nuclei can be ignored.
The Hamitonian of the system and environment is represented as
\begin{equation}\label{1}
  H_{s+e}=-\omega_{1}I_{z}^{1}-\omega_{2}I_{z}^{2}
  -\omega_{3}I_{z}^{3}
  +2\pi J_{12}I_{z}^{1}I_{z}^{2}
  +2\pi J_{23}I_{z}^{2}I_{z}^{3}
  +2\pi J_{13}I_{z}^{1}I_{z}^{3},
\end{equation}
where $\omega_{k}/2\pi(k=1,2,3)$ are the resonance frequencies of
the 3 spins, $I_{z}^{k}$ are the matrices for the z-components of
the angular momentum, and $\hbar$ is set to 1. $J_{12}$, for
example, denotes the coupling constant between spins 1 and 2.
Before the system lies in an entangled state, $^{1}H$ nucleus is
decoupled. $J_{23}$ and $J_{13}$ are averaged to zero. The system
becomes a closed system whose Hamitonian is represented as
\begin{equation}\label{2}
 H_{s}=-\omega_{1}I_{z}^{1}-\omega_{2}I_{z}^{2}+2\pi J_{12}I_{z}^{1}I_{z}^{2}.
\end{equation}
According to Ref.[11] and the experimental technique we used for
heteronuclear system [12], the pulse sequence
$[\alpha]_{x}^{2}-[grad]_{z}-[\pi/4]_{x}^{1,2}-1/4J_{12}-
[\pi]_{y}^{1,2}-1/4J_{12}-[-5\pi/6]_{y}^{1,2}-[grad]_{z}$
transforms the system from equilibrium state to the pseudo-pure
state $|\downarrow>_{1}|\downarrow>_{2}$
 represented as
$\rho_{ef}=I_{z}^{1}+I_{z}^{2}-2I_{z}^{1}I_{z}^{2}$. Here
$[\alpha]_{x}^{2}$, for example, refers to a radio-frequency(rf)
pulse on spin 2, oriented along x-axis. The evolution caused by
the pulse is denoted as $e^{i\alpha I_{x}^{2}}$. $[grad]_{z}$
denotes a gradient pulse along z-axis. $1/4J_{12}$ denotes the
evolution caused by $H_{s}$ for $1/4J_{12}$ without pulses.
Because of the difference of abundance of $^{13}C1$ and $^{13}C2$
in the sample and the effects caused by decoupling, the
equilibrium state of the system is represented as
$\gamma_{C}^{1}I_{z}^{1}+ \gamma_{C}^{2}I_{z}^{2}$, instead of
$\gamma_{C}(I_{z}^{1}+I_{z}^{2})$, where $\gamma_{C}$ is the
gyromagnetic ratio of $^{13}C$ nucleus. We call $\gamma_{C}^{1}$
and $\gamma_{C}^{2}$ "effective gyromagnetic ratios" for $^{13}C1$
and $^{13}C2$, respectively.
$\alpha=arccos(\gamma_{C}^{1}/\gamma_{C}^{2})$, where
$\gamma_{C}^{1}/\gamma_{C}^{2}$ can be measured by experiment.

The pulse sequence $[\frac{\pi}{2}]_{x}^{1,2}-
\frac{1}{4J_{12}}-[\pi]_{x}^{1,2}-\frac{1}{4J_{12}}-[\frac{\pi}{2}]_{y}^{2}$
 transforms $|\downarrow>_{1}|\downarrow>_{2}$ into an entangled
 basis state [13]
\begin{equation}\label{3}
  \rho_{s}(0)=I_{x}^{1}I_{x}^{2}-I_{z}^{1}I_{z}^{2}-I_{y}^{1}I_{y}^{2},
\end{equation}
 As soon as the system lies in the entangled
state, the decoupling pulses are closed. The system and
environment evolute under $H_{s+e}$. By applying a hard
(nonselective) pulse $[\pi]_{x}^{1,2,3}$ in the middle of a period
of evolution time t ( shown in Fig.2 ), the effect of chemical
shift evolution during this period can be cancelled [14][15]. The
system evolutes under $ H_{ef}$ represented as
\begin{equation}\label{4}
  H_{ef}=2\pi J_{12}I_{z}^{1}I_{z}^{2}+2\pi J_{23}I_{z}^{2}I_{z}^{3}+2\pi
  J_{13}I_{z}^{1}I_{z}^{3},
\end{equation}
where the last two terms describe the interaction between the
system and environment. Only these two terms cause the system to
evolute, because entangled basis states are the eigenstates of
$I_{z}^{1}I_{z}^{2}$ [16]. After t, the environment degrees of
freedom are introduced into the density matrix of the system
described as
\begin{equation}\label{5}
  \rho_{s}(t,I_{z}^{3})=cos(\varphi_{13}+\varphi_{23})(I_{x}^{1}I_{x}^{2}-I_{y}^{1}I_{y}^{2})
  -I_{z}^{1}I_{z}^{2}
  +2sin(\varphi_{13}+\varphi_{23})(I_{x}^{1}I_{y}^{2}+I_{y}^{1}I_{x}^{2})I_{z}^{3},
\end{equation}
where $\varphi_{13}=\pi J_{13}t$, and $\varphi_{23}=\pi J_{23}t$.
By tracing over the environment degrees of freedom, the density
matrix of the system is obtained, which is represented as
\begin{equation}\label{6}
\rho_{s}(t)=_{3}<\uparrow|\rho_{s}(t,I_{z}^{3})|\uparrow>_{3}
+_{3}<\downarrow|\rho_{s}(t,I_{z}^{3})|\downarrow>_{3}
=cos(\varphi_{13}+\varphi_{23})(I_{x}^{1}I_{x}^{2}-I_{y}^{1}I_{y}^{2})
  -I_{z}^{1}I_{z}^{2},
\end{equation}
using $I_{z}^{3}|\uparrow>_{3}=\frac{1}{2}|\uparrow>_{3}$, and
$I_{z}^{3}|\downarrow>_{3}=-\frac{1}{2}|\downarrow>_{3}$.
 In NMR
experiments, Eq.(6) is equivalent to the deviation density matrix
[13] represented as

\begin{equation}\label{7}
  \rho_{s}(t)=\left(\begin{array}{cccc}
    1 & 0 & 0 & -cos(\varphi_{13}+\varphi_{23}) \\
    0 & 0 & 0 & 0 \\
    0 & 0 & 0& 0 \\
    -cos(\varphi_{13}+\varphi_{23}) & 0 & 0 & 1 \
  \end{array}\right),
\end{equation}
where basis states are arrayed as $|\uparrow>_{1}|\uparrow>_{2}$,
$|\uparrow>_{1}|\downarrow>_{2}$,
$|\downarrow>_{1}|\uparrow>_{2}$,
$|\downarrow>_{1}|\downarrow>_{2}$. For convenience, we also use
$\rho_{s}(t)$ in Eq.(7). The function
$cos(\varphi_{13}+\varphi_{23})$ describes the decoherence
behavior which can be observed in the NMR spectrum through a
readout pulse $[\frac{\pi}{2}]_{x}^{2}$ which transforms the
system into the state represented as
\begin{equation}\label{8}
 \rho_{sr}(t)=\left(\begin{array}{cccc}
    0 & i & -icos(\varphi_{13}+\varphi_{23}) & cos(\varphi_{13}+\varphi_{23}) \\
    -i & 0 & cos(\varphi_{13}+\varphi_{23}) & icos(\varphi_{13}+\varphi_{23}) \\
    icos(\varphi_{13}+\varphi_{23}) & cos(\varphi_{13}+\varphi_{23}) & 0& -i \\
    cos(\varphi_{13}+\varphi_{23}) & -icos(\varphi_{13}+\varphi_{23}) & i & 0 \
  \end{array}\right).
\end{equation}

Experimental data are taken at controlled temperature (22$^{0}C$)
with a Bruker DRX 500 MHz spectrometer. The coupling constants
$J_{12}=103.1$Hz, $J_{23}=201.3$Hz, and $J_{13}=9.23$Hz. $^{1}$H
nucleus is again decoupled during recording the FID signal. In our
experiments, decoupling is the process of tracing over the
environment degrees of freedom to get the reduced density matrix
of the system, i.e., Eq.(6) is obtained from Eq.(5). Fig.3a is the
carbon spectrum through the readout pulse
$[\frac{\pi}{2}]_{x}^{2}$ when $t=3.50$ms. The center frequencies
of $^{13}C1$ and $^{13}C2$ are 124.16 ppm and 117.00 ppm,
respectively. Fig.3b shows that the amplitude of the left peak of
$^{13}C1$ varies as $t$. The experimental data points can be
fitted as function $5.8cos(2\pi
 t/T)$, where $T=8.72ms$. In theory, $T = 9.50$ms. The error
 is about $8.2\%$. It mainly results from the imperfection
of pulses, and the effect of decoherence which cannot be
controlled.

If environment is composed of multiple qubits, the interaction
between the system and environment is represented as
\begin{equation}\label{9}
  H_{i}=\sum_{k=3}^{N+2}(2\pi J_{1k}I_{z}^{1}I_{z}^{k}+2\pi
  J_{2k}I_{z}^{2}I_{z}^{k}).
\end{equation}
Under the evolution induced by $H_{i}$, the system is described as
\begin{equation}\label{10}
  \rho_{s}(t)=I_{x}^{1}I_{x}^{2}\prod_{k=3}^{N+2}cos(\varphi_{1k}+\varphi_{2k})
  -I_{z}^{1}I_{z}^{2}-I_{y}^{1}I_{y}^{2}\prod_{k=3}^{N+2}cos(\varphi_{1k}+\varphi_{2k}),
\end{equation}
through tracing over environment degrees of freedom, where
$\varphi_{1k}=\pi J_{1k}t$, and $\varphi_{2k}=\pi J_{2k}t$. If $
N\rightarrow \infty$, the coherence usually approaches to 0 very
fast,if $t\neq 0$.

In our experiments, we examine quantum decoherence of a system in
an entangled state. It is unnecessary to consider the state of
environment. In fact, we usually cannot describe the state of
environment clearly. The interaction between the system and
environment introduces the environment degrees of freedom into the
state of the system. Decoherence is the result of tracing over
such degrees of freedom.

This work was partly supported by the National Nature Science
Foundation of China. We are also grateful to Professor Shouyong
Pei of Beijing Normal University for his helpful discussions on
the theories in quantum mechanics and Miss Jinna Pan for her help
during experiments.
% ----------------------------------------------------------------------
\newpage
\bibliographystyle{article}

%-------------------------------------------------------------------------
\newpage
{\begin{center}\large{Figure Captions}\end{center}
\begin{enumerate}

\item The structure of trichloroethylene. The two $^{13}C$ nuclei and one
$^{1}H$ nucleus represent qubits 1, 2, and 3, respectively.

\item The scheme used to cancel the chemical shift evolution. $t$
denotes the evolution time. Nonselective $\pi$ pulses are applied
in the middle of $t$.

\item The experimental results when the two $^{13}$C nuclei
in an entangled state evolute under $H_{ef}$(see the text).
$^{1}$H nucleus is decoupled during recording the FID signal.
Fig.3a is the carbon spectrum through a readout pulse
$[\frac{\pi}{2}]_{x}^{2}$ when $t=3.50$ ms. The center frequencies
of $^{13}C1$ and $^{13}C2$ are 124.16 ppm and 117.00 ppm,
respectively. Fig.3b shows that the amplitude of the left peak of
$^{13}C1$ varies as t. The experimental data points can be fitted
as function $5.8cos(2\pi t /T)$, where $T=8.72ms$.
\end{enumerate}
%-----------------------------------------------------------------------

%-------------------------------
\begin{figure}{1}
\includegraphics[]{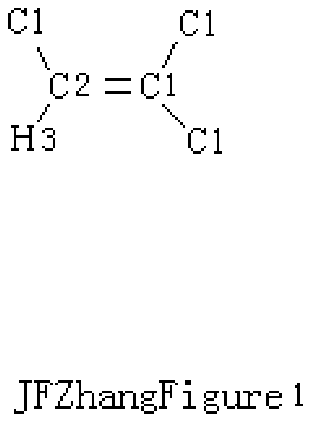}
\caption{}
\end{figure}
%-------------------------------
\begin{figure}{2}
\includegraphics[]{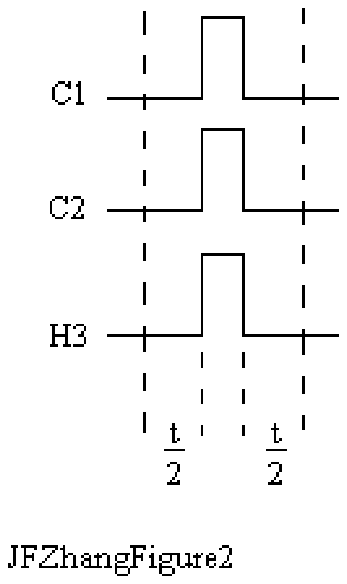}
\caption{}
\end{figure}
%-------------------------------
%-------------------------------
\begin{figure}{2}
\includegraphics[]{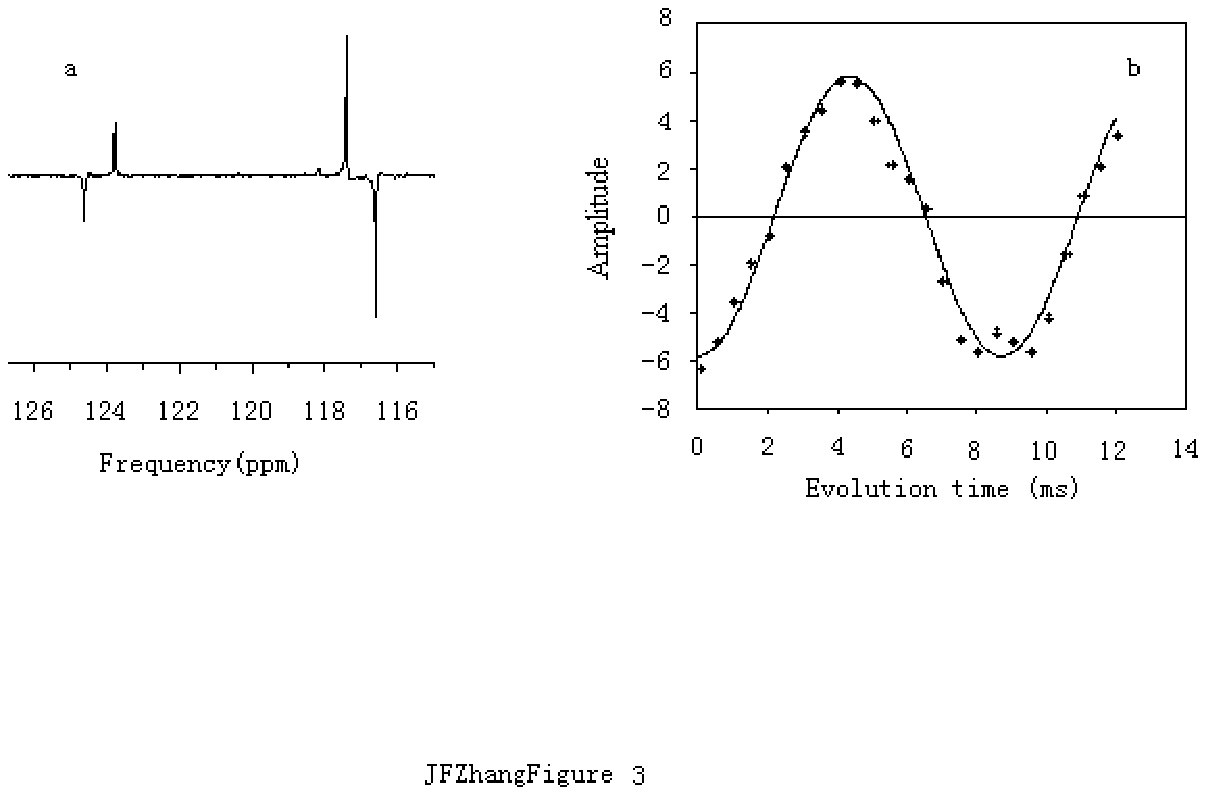}
\caption{}
\end{figure}
\end{document}